\documentclass[]{interact}
\graphicspath{{"/Users/jamesgibson/Dropbox/Aachen PostDoc/Paperwork/JARA MAX Paper/images/"}}
\usepackage{setspace,multirow,subcaption,cite}
\usepackage{lineno}
\usepackage[parfill]{parskip}
\usepackage{color, soul}

\begin{document}
\title{Mechanical Characterisation of the Protective Al$_2$O$_3$ Scale in Cr$_2$AlC MAX phases}

\author{
\name{J. S.K.-L. Gibson\textsuperscript{a}\thanks{CONTACT James S.K.-L. Gibson. Email: gibson@imm.rwth-aachen.de Tel: ‭+49 (0)241 8028298‬ Fax: +49 (0)241 8022301}, J. Gonzalez-Julian\textsuperscript{b}, S. Krishnan\textsuperscript{a}, R. Va\ss en\textsuperscript{b}  and S. Korte-Kerzel\textsuperscript{a}}
\affil{\textsuperscript{a} Institute of Physical Metallurgy and Metal Physics, RWTH Aachen University, 52074 Aachen, Germany}
\affil{\textsuperscript{b} Forschungszentrum J\"ulich GmbH, Institute of Energy and Climate Research, Materials Synthesis and Processing (IEK-1), 52425 J\"ulich, Germany}
}
\maketitle

\doublespace
\begin{abstract}
MAX phases have great potential under demands of both high-temperature and high-stress performance, with their mixed atomic bonding producing the temperature and oxidation resistance of ceramics with the mechanical resilience of metals.

Here, we measure the mechanical properties up to 980C by nanoindentation on highly dense and pure Cr$_2$AlC, as well as after oxidation with a burner rig at 1200$^{\circ}$C for more than 29 hours. Only modest reductions in both hardness and modulus up to 980$^{\circ}$C were observed, implying no change in deformation mechanism.

Furthermore, micro-cantilever fracture tests were carried out at the Cr$_2$AlC/Cr$_7$C$_3$ and Cr$_7$C$_3$/Al$_2$O$_3$ interfaces after the oxidation of the Cr$_2$AlC substrates with said burner rig. The values are typical of ceramic-ceramic interfaces, below 4~MPa$\sqrt{m}$, leading to the hypothesis that the excellent macroscopic behaviour is due to a combination of low internal strain due to the match in thermal expansion coefficient as well as the convoluted interface. 
\\
\end{abstract}



\section{Introduction}
There is a strong demand for structural materials to work under the increasingly harsh condition of high temperatures, oxidising and corrosive environments and high stresses, particularly in the energy and transport industries. Power generation plants such as stationary and mobile gas turbines, concentrated solar units or heat exchangers are some examples of systems that will clearly benefit from the replacement of the current metal or superalloy components with new materials with improved properties. In that sense, MAX phases have attracted great interest in recent years due to the unique combination of properties, bridging the gap between ceramics and metals~\cite{Barsoum1999363}. MAX phases have the general formula of M$_{n+1}$AX$_n$, where M corresponds to an earlier transition metal, A is a A-group element (IIIA or IVA element), X is C or N, and $n$ is typically equal to 1, 2 or 3~\cite{Eklund20101851}. In general, this relatively unused class of materials are lightweight, present high elastic modulus, and are resistant to chemical attack, while retaining good machinability, thermal shock resistance, damage tolerance and high electrical and thermal conductivity~\cite{Barsoum2011195,Tian20071663,Sun2011143}. In terms of high temperature applications, those MAX phases that contain Al as “A” element are the most attractive due to the in-situ formation of an external and protective $\alpha$-A$_2$O$_3$ layer~\cite{Farle201537,Li2013892}. In particular, the most interesting and studied MAX phases are Ti$_2$AlC, Ti$_3$AlC$_2$ and Cr$_2$AlC~\cite{Sun2011143,Barsoum2011195,Li2013892,Wang2010385,Walter2006389,Shtansky20093595,Horlait2016682}.

The ternary carbide Cr$_2$AlC is one of the most promising MAX phases for high temperature applications due to its excellent combination of properties. Cr$_2$AlC presents a density of 5.23 g/cm$^3$, high Young's modulus (288~GPa), intermediate fracture toughness (5.2~MPa$\sqrt{m}$), good flexural (354~MPa) and compression strengths (1160~MPa)~\cite{Barsoum2011195,Yu20105997,Tian20071663}. Furthermore, bulk Cr$_2$AlC materials are good electrical and thermal conductors, show excellent oxidation resistance at high temperature in air and exhibit self-healing mechanisms at high temperature~\cite{Berger2017192,Pei2017441,Shen20161}. However, their high-temperature mechanical properties are largely unknown, and the mechanism responsible for the oxidation resistance is not fully understood.

Cr$_2$AlC has been investigated by several authors concerning a variety of its properties under potential application conditions. Gupta measured the tribological behaviour of a variety of MAX phases against alumina~\cite{Gupta2008560} and a Ni-superalloy~\cite{Gupta2008270}, demonstrating for Cr$_2$AlC a relatively slow sliding friction ($\mu \sim 0.4$ and $\sim 0.6$, respectively), but a very noisy contact with alumina due to the non-uniform formation of the tribofilm. Most research however has focussed on the oxidative properties of the material. Li and coworkers studied the influence of grain size on the oxidation behaviour of Cr$_2$AlC at 1100$^{\circ}$C and 1200$^{\circ}$C~\cite{Li20132715}. As would be expected, fine grain sizes and higher temperatures increased oxidation rates, and it was observed that the post-oxidisation microstructure consisted of an Al$_2$O$_3$ scale on top of a Cr$_7$C$_3$ layer covering the MAX phase. This oxidation behaviour - in particular the formation of Al$_2$O$_3$ and Cr$_7$C$_3$ layers - was also observed by Shang et al.~\cite{Shang20171339}, as well as Berger et al.~\cite{Berger2015373} for magnetron-deposited Cr$_2$AlC films. Yang et al.~\cite{Yang2013203} attribute this to an outward growth of the alumina scale via the diffusion of aluminium ions via grain boundaries, followed by the decomposition of the Cr$_2$AlC to form the Cr$_7$C$_3$ layer. It is this alumina-formation that has led to the interest of several groups in the potential self-healing properties of Cr$_2$AlC~\cite{Shen20161,Berger2017192,Pei2017441}.

These studies have usually been performed under controlled conditions, using for example thermogravimetric analysis (TGA). The oxidation response is highly controlled by the purity of the MAX phases - indeed synthesis is one of the main difficulties of these materials - but after several investigations it has been concluded that the oxidation kinetics are described by cubic kinetics in the range of temperatures between 900 - 1400$^{\circ}$C in air~\cite{Hajas2011591,Li2013892,Tallman2013115}. Nevertheless, the final components will operate under different environmental conditions such as thermal gradients, rapid heating and cooling (thermal shock), water vapour (critical for advanced ceramics as SiC), and other corrosive chemicals. Furthermore, in most of these applications the component has to withstand these aggressive environments for at least thousands of hours, so thermal barrier coatings (TBCs) are typically required. These two points - oxidation response at high temperature under realistic conditions and interaction with TBCs - are determinant to transfer the Al-based MAX materials into final applications. Unfortunately, only a few works have been published in that direction~\cite{Smialek201723,Gonzalez20181841,Gonzalez201817,Smialek201677,Smialek2018782}, despite the fact that preliminary results are highly promising. Highly pure and dense Cr$_2$AlC materials have been recently tested under realistic environmental gas turbine conditions using a burner rig facility~\cite{Gonzalez201817}. The sample survived perfectly after more than 58 hours under 500 cycles at 1200$^{\circ}$C due to the formation of a highly dense, well-adhered and protective $\alpha$-Al$_2$O$_3$ layer. Similar results were reported by Smialek et al.~\cite{Smialek201723} in dense Ti$_2$AlC materials. The cubic rate constants were approximately 20\% of those measured in static conditions, with a small recession rate of 0.012 mg$\cdot$cm$^{-2}$$\cdot$h at 1300$^{\circ}$C using a high-pressure burner rig operated at 6 atm, 25 m/s, and $\sim$10\% water vapour. Furthermore, the interaction of these MAX phases with TBCs has been also tested at high temperature for long times, showing an excellent performance due to the coefficient of thermal expansion (CTE) match between the MAX phase, the formed $\alpha$-A$_2$O$_3$ layer and the YSZ used as TBC material~\cite{Gonzalez201817,Smialek201677}.

Therefore, the success of these Al-based MAX phases and their potential transfer to industrial applications mainly depend on two factors: i) formation and oxidation kinetics of the oxide scale, and ii) mechanical interlocking between the MAX phase and the oxide scale, especially under thermal shock conditions. As previously mentioned, the formation and the oxidation kinetics have been widely studied. However, to the best of our knowledge, the mechanical properties and/or the adhesion between the different layers have not been characterized. There are very little data on the mechanical properties of the material at the intended operating temperatures, as well as the mechanical properties of the critically-important alumina and chromium carbide layers. Due to the recent advances in nano-mechanical testing, it is now possible to test these materials near the operating temperatures of the components~\cite{Gibson2017007,Korte2012167,Harris20161}, 

These MAX phase materials have been investigated in this work by high-temperature nanoindentation and micro-cantilever bending to measure the mechanical properties of the Al$_2$O$_3$ and Cr$_7$C$_3$ as well as the fracture toughness of the interfaces.   

\section{Materials and Methods}
\subsection{Sample Preparation}
Single-phase, dense Cr$_2$AlC samples were produced in a two-step process, as described elsewhere~\cite{Gonzalez2016415}. Briefly, the elemental constituents were mixed, followed by a thermal treatment at 1400$^{\circ}$C in vacuum for three hours to synthesize the Cr$_2$AlC phase. Then, the pellets were ground and milled to obtain the powder with a unimodal particle size distribution and a mean particle size around 9~$\mu$m. Afterwards, this powder was fully densified using a Field Assisted Sintering Technology/Spark Plasma Sintering (FAST/SPS, FCT-HPD5, FCT Systeme GmbH, Germany). The sintering conditions were heating rate of 100 K/min, maximal temperature of 1300$^{\circ}$C, isothermal holding time of five minutes, and uniaxial pressure of 50~MPa during the whole thermal cycle. The highly pure Cr$_2$AlC samples were discs of 30~mm diameter and 5~mm height. One of these samples was then taken as-is and prepared to a 1~$\mu$m diamond finish for nanomechanical testing.

Oxidation tests using a second of these samples was carried out using a burner rig as described in~\cite{Gonzalez20181841}. Briefly, the gas burner rig was operated with natural gas and oxygen in a ratio 1:2.2. The surface temperature was fixed at 1200$^{\circ}$C using a long-wave pyrometer, and 500 thermal cycles were continuously performed. One thermal cycle consists of five minutes of heating by the impact of a flame plume on the sample surface, followed by two minutes of cooling due to the lateral displacement of the burner. The sample was tested under these conditions for 3500 minutes (58 hours and 20 minutes), with 1750 minutes (29 hours and 10 minutes) at 1200$^{\circ}$C. It was then sectioned parallel to the incident direction of the flame, such that the oxidized surface layers were visible, and then prepared to a 1~$\mu$m diamond finish. The two samples - one that was not exposed to the burner rig, and one that was subjected to the above conditions - will be referred to as the `standard' and `flame' samples, respectively.

\hl{In both samples, the mean particle size of the Cr$_2$AlC is $5.1 \pm 0.2$~$\mu$m as previously reported}~\cite{Gonzalez20181841}\hl{, and no texture is expected due to the  combination of the initial random grain orientation with little grain growth during densification due to the SPS process.}

\subsection{Nanoindentation}
Room temperature nanoindentation was performed using a Nanomechanics iNano system equipped with a diamond Berkovich indenter at a constant strain rate of 0.2~s$^{-1}$ to analyse the mechanical properties of the individual phases in the flame sample, with the final data taken as an average between 150-300~nm. Additional nanoindentation was performed on the standard sample using a MicroMaterials (MML) NanoTest Platform-3 system situated in a custom-built vacuum chamber ($\sim$$10^{-5}$~mbar). Indents were performed to 1000~nm using an independently-heated Berkovich sapphire tip. Tests were conducted at a constant load rate of 20~mN/s, followed by a 5 second dwell period and 20~mN/s unloading rate. Up to and including 700$^{\circ}$C, 50 indents were carried out at each temperature. Due to the extremely high rates of tip wear at very high temperatures, only six indents were performed at 980$^{\circ}$C. \hl{The flame sample was only investigated using the iNano, and the standard sample only investigated by the MicroMaterials indenter}, but on both instruments, the area function of the tip and frame stiffness were determined prior to the start of indentation by indents on fused silica according to the method of Oliver and Pharr~\cite{Oliver19921564,Oliver20043}.

\subsection{Fracture Toughness}
Micro-cantilevers were milled on the flame sample using an FEI Helios 600i focussed ion beam (FIB), based on the method of Di Maio and Roberts~\cite{Maio2005299}. This method, producing cantilevers with a pentagonal cross-section, was chosen as it allows the targeting of specific boundaries of interest, without the need for these to be on the edge of the sample which would be the case for a simplified rectangular geometry. Suitable sites in this case were those such that the sites contain a straight inter-phase boundary and were free of significant porosity or micro-cracking. This was achieved using BSE (backscattered electron) imaging to clearly identify the different phases. The beams had an approximate length of 10~$\mu m$, with a width and depth of 3~$\mu m$. Initial cuts were made at 21~nA, with the beams subjected to a final `polishing' beam current of 0.23~nA. Notches were made at the grain boundary at the lowest available current of 1.1~pA, with all milling taking place at 30~kV. Three beams were manufactured and tested at the Al$_2$O$_3$-Cr$_7$C$_3$ interface, and an additional beam at the Cr$_7$C$_3$-Cr$_2$AlC interface. Testing was carried out at room temperature \textit{in-situ} using a Nanomechanics InSEM-III inside a Tescan Vega-3 scanning electron microscope (SEM). A cube corner indenter was used to displace the free end of the beam at a constant load rate of 0.1~mN/s until failure was observed. Fracture toughness was calculated using the stress intensity factor and geometry function as determined by Di~Maio and Roberts~\cite{Maio2005299}.

\section{Results}
\begin{figure}[h]
    \centering
     \includegraphics[width=.9\linewidth]{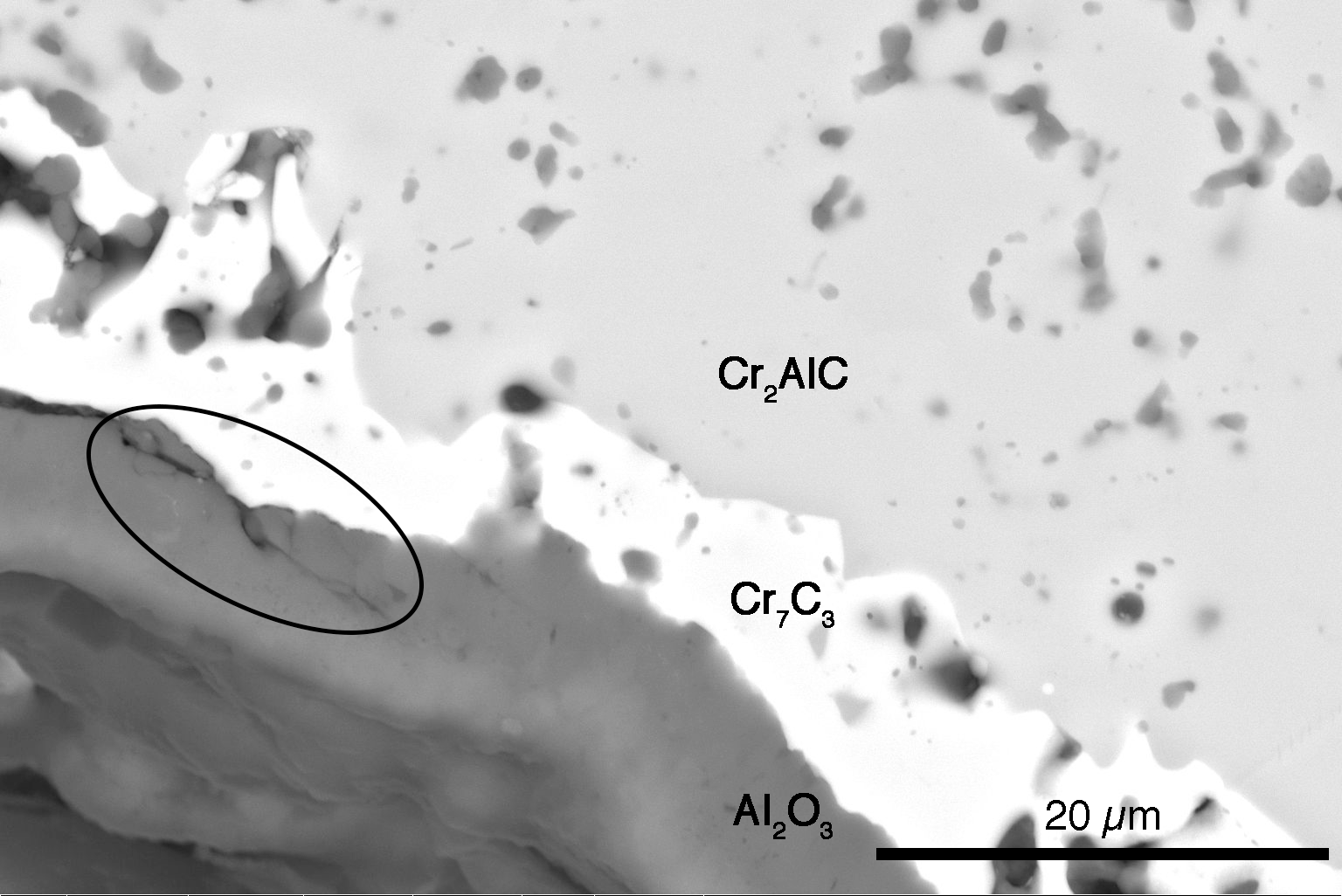}
     \caption{Microstructure of the flame sample. Approximately 10~$\mu$m thick layers of Cr$_7$C$_3$ and Al$_2$O$_3$ have formed on the surface as a result of the cyclic burner tests, with the Cr$_7$C$_3$ layer in particular showing a large degree of porosity. \hl{Cracking in the alumina layer is shown circled. Some secondary phases can be seen in the Cr$_2$AlC, likely alumina (in dark) and some smaller carbide and aluminides (light).}}
    \label{Microstructure}
\end{figure}

The microstructure of the sample is shown in Figure~\ref{Microstructure}, consisting of bulk Cr$_2$AlC with layers of Cr$_7$C$_3$ and Al$_2$O$_3$ on top. The Cr$_7$C$_3$ layer shows a significant degree of porosity, and the layers are non-planar and non-parallel, as a result of the oxidation of a powder-processed sample. \hl{Some secondary phases can be seen in the Cr$_2$AlC, likely alumina (in dark) and some smaller carbide and aluminides (light), although this was not possible to confirm by EDX due to their small size. The porosity of the Cr$_2$AlC is around 1\%, as reported in previous work}~\cite{Gonzalez20181841}.

\subsection{Nanoindentation}
\begin{figure}[ht]
	\centering
	\includegraphics[width=.7\linewidth]{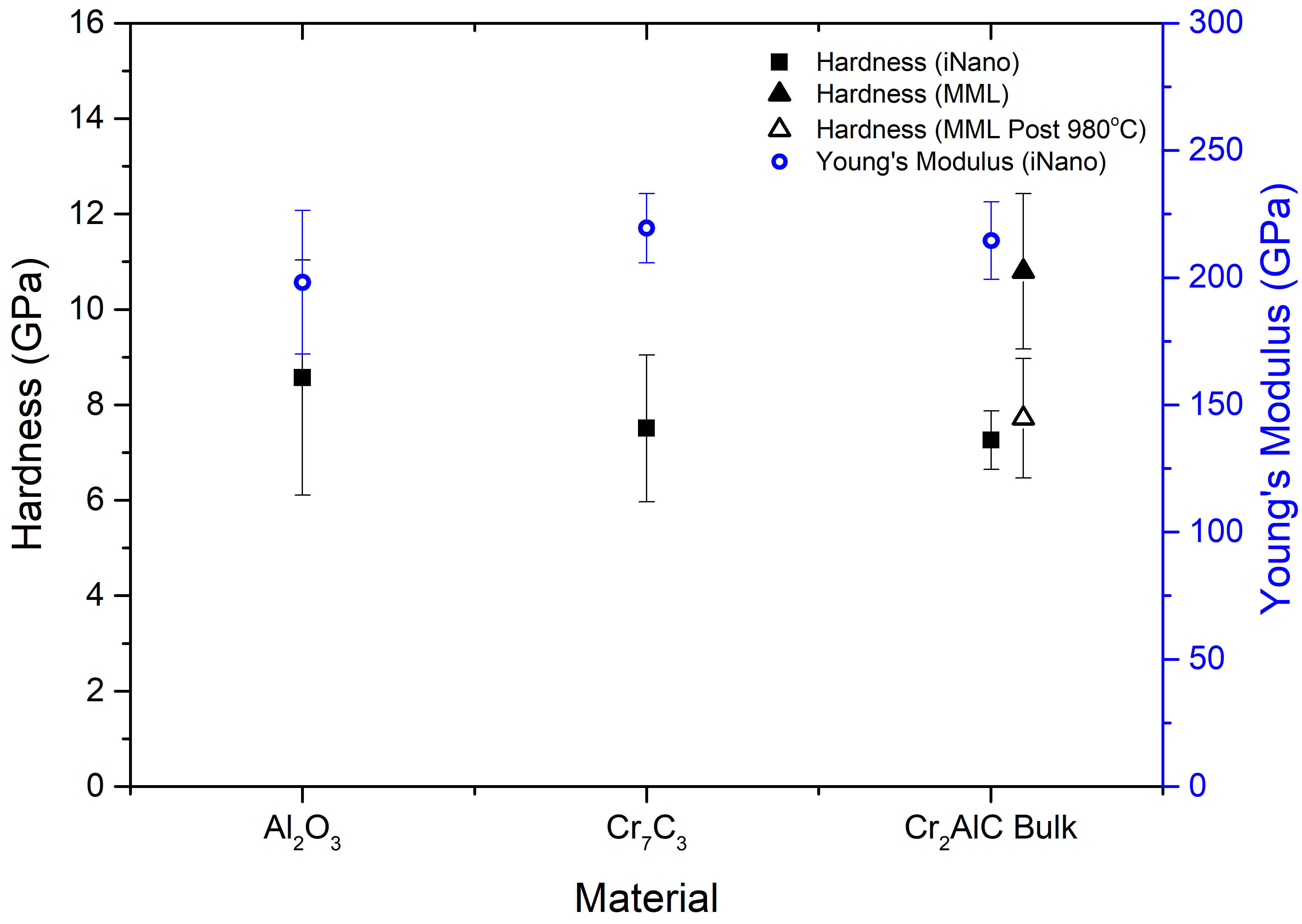}
	\caption{Hardness and elastic modulus of the three layers found at the surface of the flame sample. Error bars indicate one standard deviation in the scatter of the results. The initial hardness and the post-980$^{\circ}$C hardness as measured by the MML high-temperature system are included for comparison.}
	\label{RoomTNano}
\end{figure}

Figure~\ref{RoomTNano} shows the room-temperature nanoindentation data, namely the hardness and Young's modulus of the Al$_2$O$_3$ and Cr$_7$C$_3$ layers, as well as the Cr$_2$AlC bulk. The Cr$_2$AlC data is consistent between the two indenters - particularly the post-980$^{\circ}$C hardness - despite the difference in indentation depth and analysis method (CSM vs an Oliver-Pharr fit). A similar hardness and elastic modulus are seen for all three phases, with the Al$_2$O$_3$ slightly softer and more compliant, albeit with slightly higher spread in the data, likely due to porosity or micro-cracks in this layer.

\begin{figure}[ht]
    \centering
    \begin{subfigure}[t]{0.5\textwidth}
        \centering
        \includegraphics[width=.9\linewidth]{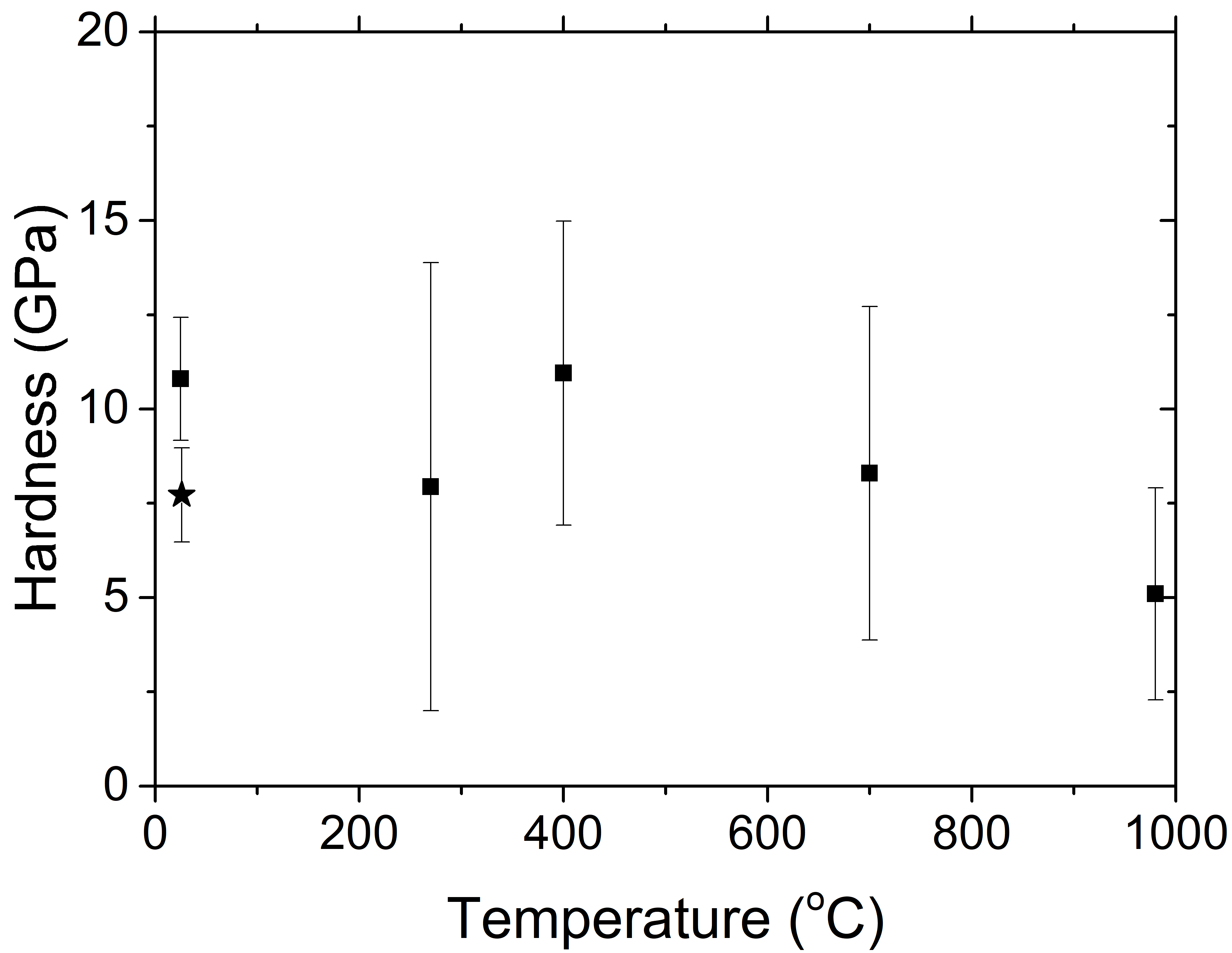}
        \caption{}\label{H_1000_AVG}
    \end{subfigure}%
    ~ 
    \begin{subfigure}[t]{0.5\textwidth}
        \centering
        \includegraphics[width=.9\linewidth]{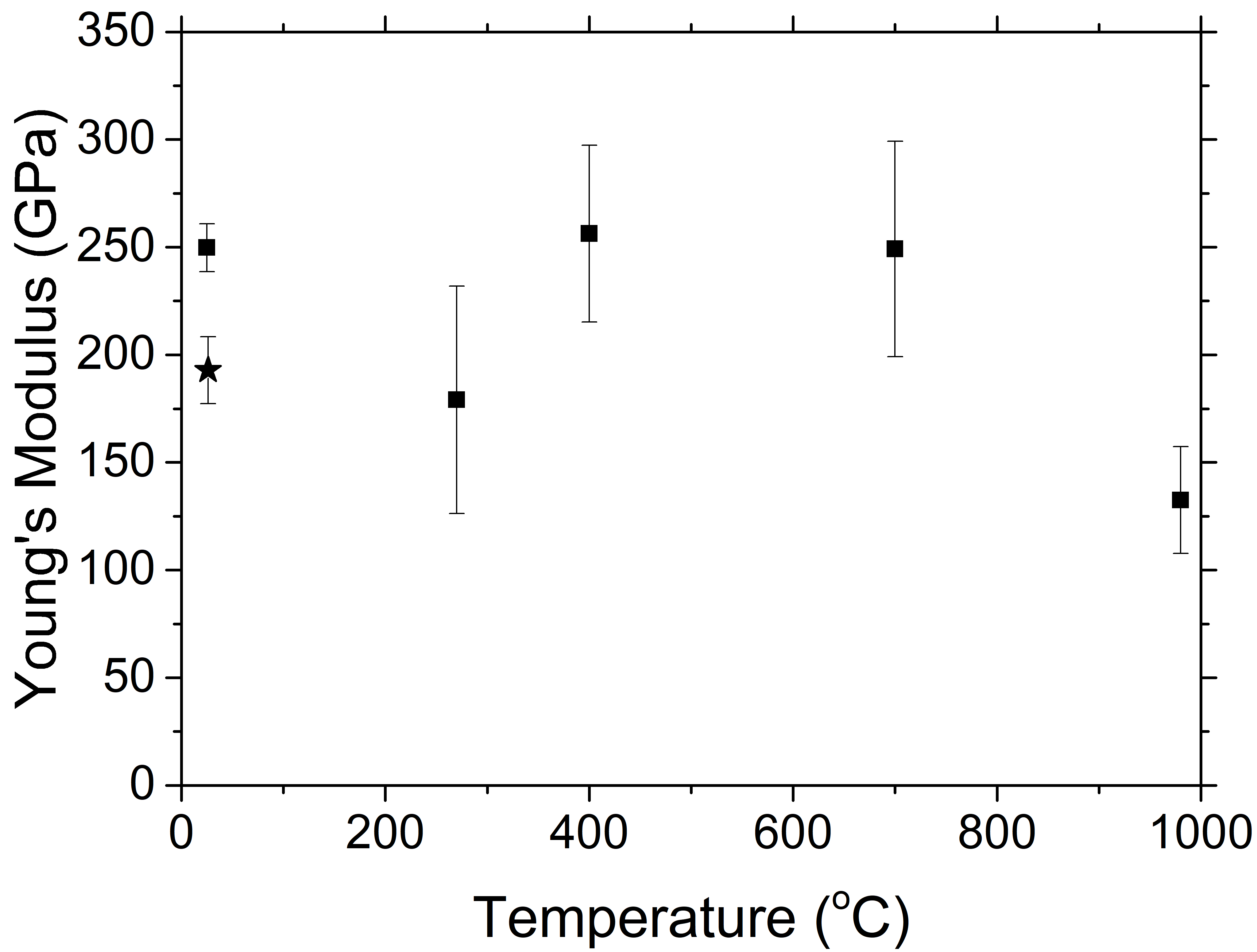}
        \caption{}\label{E_1000_AVG}
    \end{subfigure}
    \caption{(\textbf{a}) Indentation hardness with temperature for Cr$_2$AlC at 1000~nm. (\textbf{b}) Young's modulus with temperature for Cr$_2$AlC up to 980$^{\circ}$C. In both cases, the starred point at room temperature represents data collected after the sample had been heated to 980$^{\circ}$C. Error bars indicate one standard deviation in the scatter of the results, i.e.~the displayed bar covers two standard deviations in total. \hl{All tests were carried out on the flame sample, i.e. the sample has been previously exposed to the 1200$^{\circ}$C burner rig.}}
    \label{NanoResults}
\end{figure}

\begin{figure}[ht]
    \centering
    \includegraphics[width=.7\linewidth]{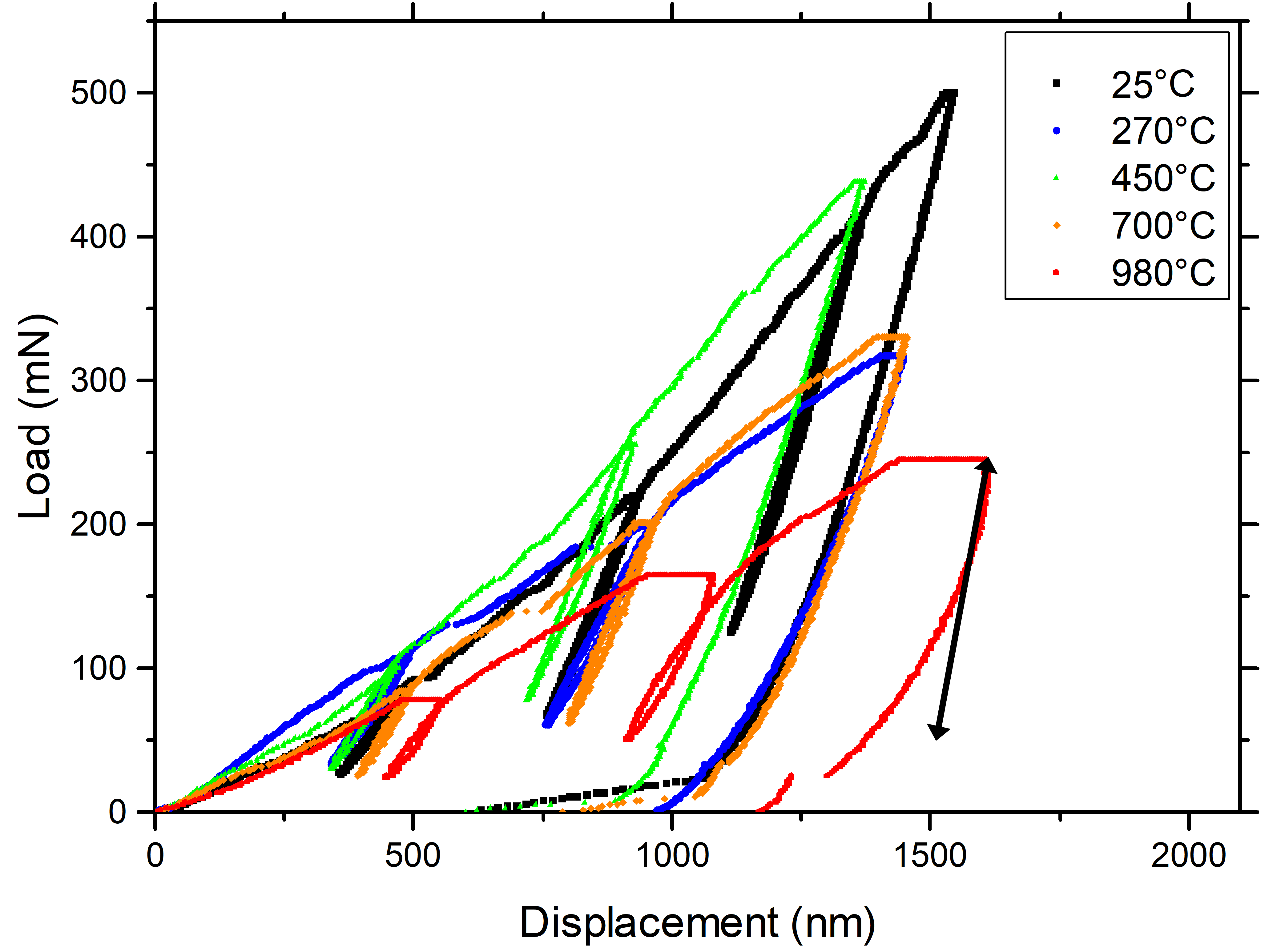}
    \caption{Load-Displacement curves for an average indent (i.e.~an indent producing the average value of hardness and modulus) for each of the temperatures measured at elevated temperature \hl{(i.e.~on the `standard' sample)}. The black arrow next to the curve recorded at 980$^{\circ}$C shows the range the Oliver-Pharr fit to the data is applied over, namely 100-20\% of the maximum load. The non-linearity of the 980$^{\circ}$C unloading data is clearly visible.}
    \label{Hysteresis}
\end{figure}

Figure~\ref{NanoResults} subsequently displays the hardness and Young's modulus as determined by high-temperature nanoindentation. As has been done previously at very high indentation temperatures~\cite{Gibson2017007}, the temperature-dependent Young's modulus and Poisson's ratio of the indenter were used in the conversion of reduced to Young's modulus. An example indentation load-displacement curve is given in Figure~\ref{Hysteresis} for each test temperature. Up to and including 700$^{\circ}$C, the five second hold period at the peak of each loading cycle can be observed to give very little increase in depth, i.e.~almost no creep is occurring at these temperatures. At 980$^{\circ}$C, the `1000~nm' indentation depth contains approximately 125~nm of creep deformation. This creep behaviour is also evident in the unloading portion of the hysteresis curve, which is therefore likely responsible for the low value of modulus at 980$^{\circ}$C evident in Figure~\ref{E_1000_AVG}. This is due to the inaccuracy of a standard Oliver-Pharr fit~\cite{Oliver19921564} to the unloading data, as this fit assumes an entirely elastic unloading process. The black arrow next to the curve recorded at 980$^{\circ}$C shows the range of the fit, namely 100-20\% of the maximum load.

\clearpage

\subsection{Fracture Toughness}
\begin{figure}[ht]
    \centering
    \begin{subfigure}[t]{0.5\textwidth}
        \centering
        \includegraphics[width=.9\linewidth]{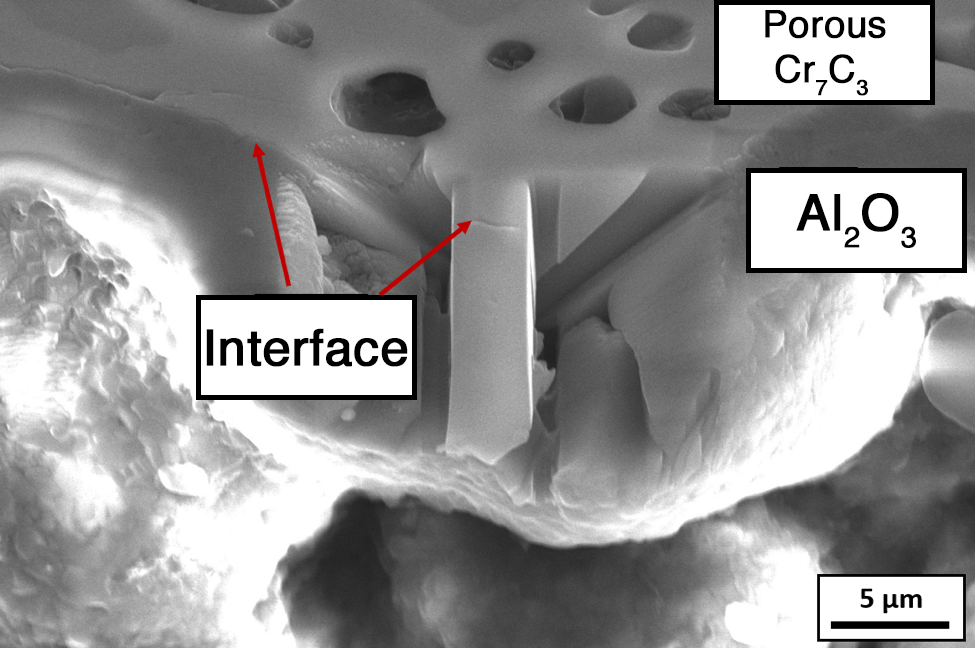}
        \caption{}\label{Cantilever}
    \end{subfigure}%
    ~ 
    \begin{subfigure}[t]{0.5\textwidth}
        \centering
        \includegraphics[width=.9\linewidth]{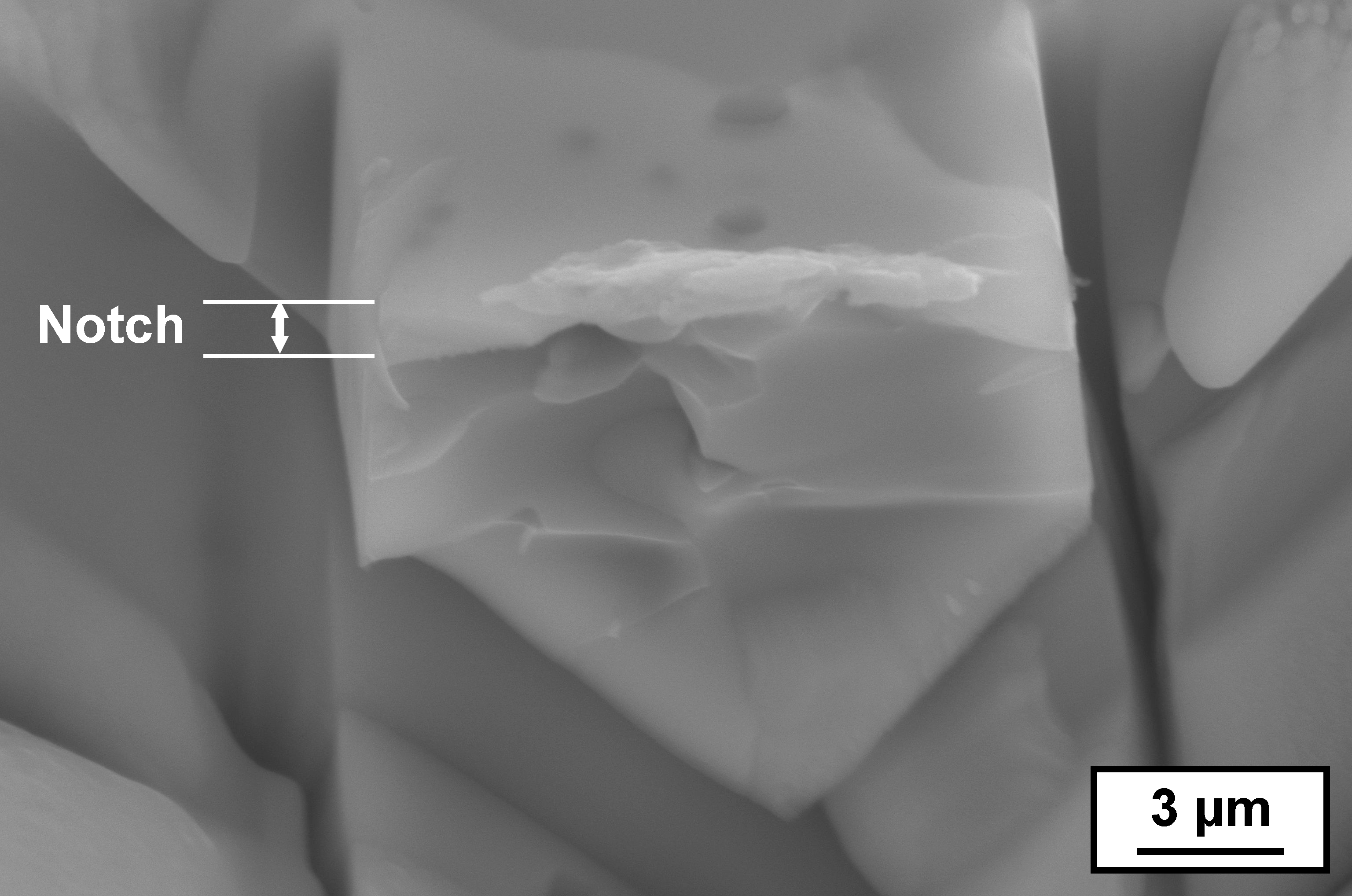}
        \caption{}\label{CantileverPost}
    \end{subfigure}%
    ~\\
    \begin{subfigure}[t]{0.5\textwidth}
        \centering
        \includegraphics[width=.9\linewidth]{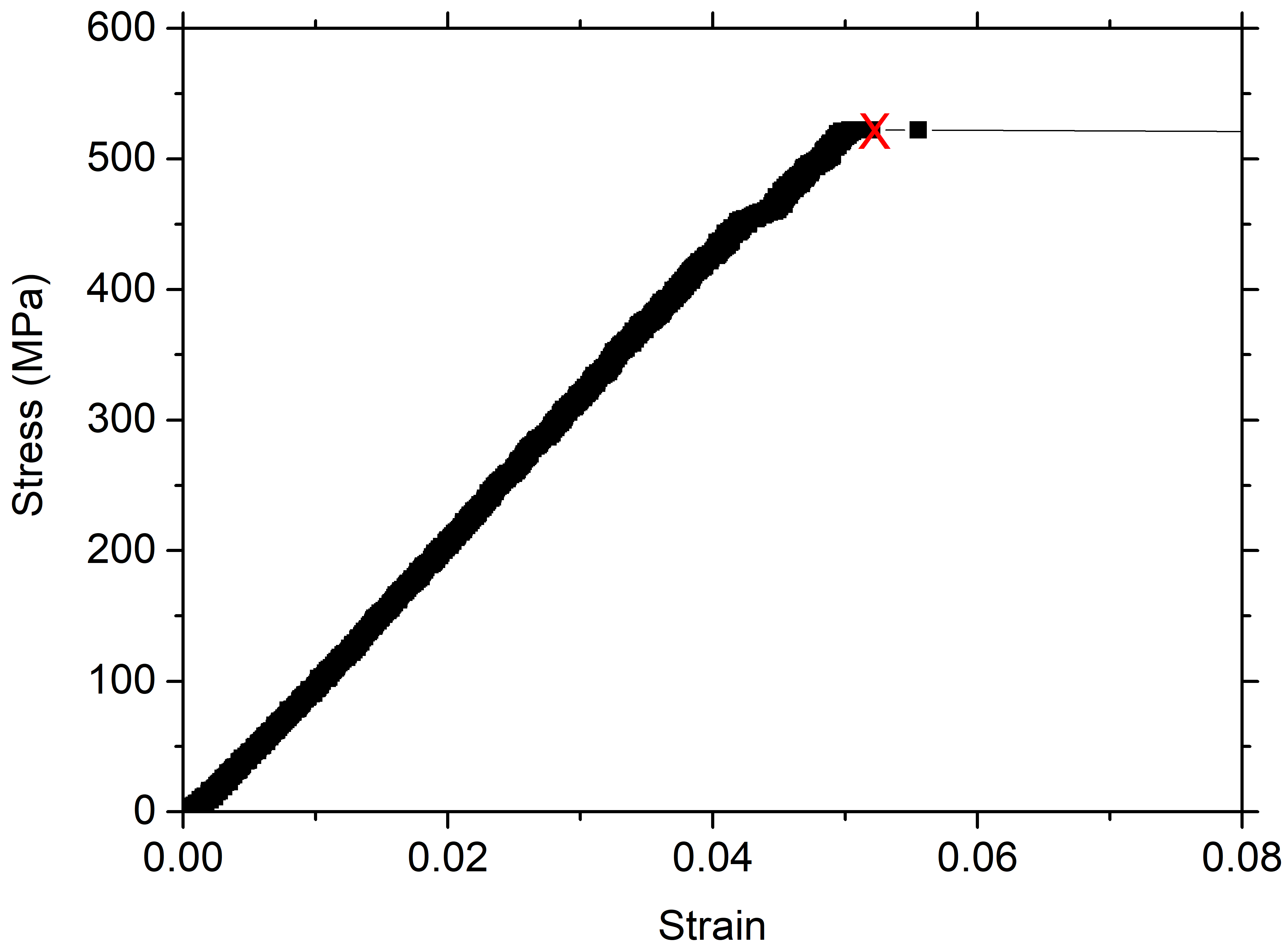}
        \caption{}\label{Beam1}
    \end{subfigure}
    \caption{(\textbf{a}) Typical micro-cantilever produced for fracture toughness tests, with a FIB-made pre-notch at the boundary of interest (here, the Cr$_7$C$_3$-Cr$_2$AlC boundary near the base of the beam). (\textbf{b}) Fracture surface of the cantilever after testing, showing intergranular failure emanating from the FIB-milled notch. (\textbf{c}) Typical stress-strain curve of a microcantilever, showing elastic loading followed by rapid, brittle fracture around 5\% strain, marked by the red `X'. The data after this point are generated automatically by the indenter travel into the trench of the cantilever.}
    \label{CantileverExample}
\end{figure}

A typical stress-strain curve for the micro-cantilevers is shown in Figure~\ref{Beam1}, where the linear nature suggests a very brittle fracture process, as would be expected at ceramic-ceramic interfaces. The fracture toughness measurements determine an average toughness of $1.62 \pm 0.8$~MPa$\sqrt{m}$ for the Al$_2$O$_3$-Cr$_7$C$_3$ interface, and 3.55~MPa$\sqrt{m}$ for the Cr$_7$C$_3$-Cr$_2$AlC interface. In all the tested beams, the propagating crack followed the interface, resulting in fracture surfaces similar to that shown in Figure~\ref{CantileverPost}.

\begin{table}[ht]
\begin{center}
\begin{tabular}{c c}
\hline
Interface & Fracture Toughness (MPa$\sqrt{m}$)\\
\hline
\multirow{3}{*}{Al$_2$O$_3$-Cr$_7$C$_3$} & 0.68 \\
 & 2.32 \\
 & 1.87 \\
\hline
\multirow{1}{*}{Cr$_7$C$_3$-Cr$_2$AlC} & 3.55 \\
\hline
\end{tabular}
\caption{Fracture toughness of the microcantilevers.
\label{KcTable}}
\end{center}
\end{table}

\section{Discussion}
Comparing the measured mechanical data to that found in the literature, Naveed et al.~\cite{Naveed201611} measured $\sim$12~GPa and $\sim$290~GPa as hardness and modulus, respectively, for sputtered Cr$_2$AlC films by nanoindentation, almost identical to our room temperature data. Also of note is that the significant error bars seen here are also present in both the hardness and the modulus of the sputtered films, due to the significant anisotropy in the mechanical behaviour of MAX phases, with dislocations only occurring in the basal plane leading to kink banding~\cite{Barsoum2011195}. Similar anisotropy in the elastic behaviour has been calculated via DFT by Sun et al.~\cite{Sun2004589}. \hl{Due to the relatively large number of indents combined with the fine grain size, it is expected that the indents cover a full range of crystal orientations and these error bars are thus representative of the full range of mechanical properties with orientation.}

Data on chromium carbide is relatively sparse in the literature, but Hou et al.~\cite{Hou20051039} measure magnetron-sputtered chromium carbide films with a hardness and elastic modulus of 17.0~GPa and 245~GPa, respectively. However, magnetron-sputtered Cr$_3$C$_2$ measured by Paul and coworkers~\cite{Paul2002123} showed a hardness between 5 and 7.5~GPa, depending on radio frequency (RF) power, and a modulus close to 175~GPa.  Xiao\cite{Xiao2009115415} calculated by DFT for Cr$_7$C$_3$  a modulus of 374~GPa. Finally, data for Cr$_3$C$_2$ given by Hussainova~\cite{Hussainova2006} states a modulus of 370 GPa. Overall, the data measured here of a hardness of 7.5~GPa and a modulus of 220~GPa can therefore be stated to agree reasonably well.

A potential discrepancy is seen in the data obtained for Al$_2$O$_3$. Here, we measure 8.6~GPa and 198~GPa as hardness and modulus, respectively. Auerkari~\cite{Auerkari1996} states a Young's modulus of 410 GPa for porosity-free material (grade A1), well above that measured here, although this drops rapidly to 260~GPa with 6\% porosity (grade A9). Hardness shows a similar variation from $\sim$20~GPa to 7.8~GPa for these two grades. The low values measured here with respect to the pure material may therefore simply be a consequence of a porous or micro-cracked surface layer \hl{- indeed cracking is visible in Figure}~\ref{Microstructure}\hl{, shown circled. As the elastic strain field around the indenter is very large, the measurement of elastic modulus comes from a hemisphere of radius around 100 times the indentation depth (i.e.~30~$\mu$m in this case). Therefore the degree of cracking or porosity need not be large to influence these measurements.}

High-temperature data for MAX phases are sparse, but Lapauw et al. measured the elastic modulus of Nb$_4$AlC$_3$ and (Nb$_{0.85}$,Zr$_{0.15}$)$_4$AlC$_3$, and showed that both materials exhibited a drop of elastic modulus to 72.4\% of their room-temperature value at 1000$^{\circ}$C~\cite{Lapauw20165445}. First-principles calculations (DFT) have been carried out for the V$_2$AlC and V$_2$GaC phases, calculating a reduction in bulk modulus from 160~GPa at room temperature to 133.5~GPa at 700$^{\circ}$C~\cite{Khatun20173634} (i.e.~71\% of RT at 1000$^{\circ}$C). Similar first-principles calculations were carried out on Ti$_3$AlC$_2$ to calculate a drop of bulk modulus from 150~GPa to 135~GPa over the same temperature range~\cite{Ali20124221} (i.e.~83\% of RT at 1000$^{\circ}$C). Here, a drop from 250~GPa initially at room temperature to 132~GPa at 980$^{\circ}$C is measured, i.e.~53\% of the room-temperature value, slightly more than these other works. This may be either due to the overestimation of theoretical calculations compared to experimental data for Cr$_2$AlC~\cite{Barsoum2011195}, or due to errors in the Oliver-Pharr fit to the unloading data given there is creep deformation still present in the unloading curve (Figure~\ref{Hysteresis}) that has been previously shown to result in low values of Young's modulus in high-temperature nanoindentation~\cite{Gibson2017007}. As mentioned, this fit assumes an entirely elastic unloading process, while the black arrow next to the curve clearly demonstrated this is not the case. This latter reason - creep deformation - seems most likely to explain the low modulus, as the drop is only significant at 980$^{\circ}$C, at the point where creep is observable in the load-displacement data. 

The mechanism of plastic deformation in MAX phases is widely attributed to dislocation motion on the basal plane and the formation of kink bands~\cite{Barsoum1999363,Barsoum2011195}, even up to very high strain rates~\cite{Bhattacharya2014319}. However, there is a change in deformation mechanism - a brittle-plastic transition - generally said to be around 1000-1100$^{\circ}$C~\cite{Barsoum2011195}, though this is likely dependent of the specific MAX phase. TEM observations showed that Ti$_2$AlN was plastically deformed at 900$^{\circ}$C under a 350~MPa confining pressure producing non-basal and cross-slipping dislocations~\cite{Guitton20146358}. In Cr$_2$AlC, the nanoindentation hardness at 980$^{\circ}$C does not significantly deviate from the linear reduction in hardness with temperature, implying there is no change of deformation mechanism in Cr$_2$AlC at temperatures up to 980$^{\circ}$C.

The hardness and modulus of the Cr$_2$AlC when re-tested at room temperature was potentially seen to drop after the tests at 980$^{\circ}$C (Figure~\ref{NanoResults}). \hl{For both hardness and modulus, the two datasets overlap, i.e. not all "starred" values are lower the lowest RT value}. The sample itself had already been exposed to the $\sim$1200$^{\circ}$C of the burner rig implying no microstructural changes have taken place. Any errors due to unaccounted-for tip blunting would over-estimate hardness or modulus, and similarly any additional densification after processing should also increase modulus. Therefore, given the standard deviation in results for much of the remaining data ($\sim$4~GPa in hardness and 50~GPa in modulus)\hl{, and the overlapping data, }it is most likely that this drop is simply an artefact of measurement location where lower-modulus grains were present. \hl{This too is likely the explanation for similar discrepancies, such as the low modulus at 300$^{\circ}$C. As a comparison, for another highly anisotropic material with a similar spread of hardness and modulus values  - WC - 90,000 indents were used to describe its deformation behaviour using} nanoindentation~\cite{ROA2018211}\hl{. However, wear and chemical interactions unfortunately currently place these large-scale maps out of the reach of indentation as these temperatures.}

Macroscopically, Cr$_2$AlC was observed to have excellent thermal stability, particularly in regards to the burner rig in which no spallation of oxide layers were seen even after 500 heating-cooling cycles~\cite{Gonzalez20181841}. The micro-mechanical tests at the interfaces confirm that despite the harsh conditions to which the sample was exposed, there is very likely no degradation of these interfaces, as they show typical values for fracture toughness of ceramic materials. Here, the fracture toughness measurements determine an average toughness of 1.62~MPa$\sqrt{m}$ for the Al$_2$O$_3$-Cr$_7$C$_3$ interface, and 3.55~MPa$\sqrt{m}$ for the Cr$_7$C$_3$-Cr$_2$AlC interface. In comparison, single crystal alumina when measured by indentation cracking~\cite{Wang20182073} is around 1.5~MPa$\sqrt{m}$, very similar to that of the alumina-based interface. Hirota measures, in a 97\% Cr$_7$C$_3$ - 3\% Cr$_2$O$_3$ sample, a fracture toughness of 6.3 MPa$\sqrt{m}$, and allowing for a broader comparison, the fracture toughness of an alumina-chromium carbide (Cr$_3$C$_2$) nanocomposite has been measured to be 5.5~MPa$\sqrt{m}$~\cite{Huang201107}. Both of these data are above, but not too far from the 3.55~MPa$\sqrt{m}$ measured at the Cr$_7$C$_3$-containing interface. While these values of fracture toughness are somewhat lower than that of the bulk Cr$_2$AlC ($\sim$6.2~MPa$\sqrt{m}$)~\cite{Yu20105997}, it is promising to see that without any design or improvement of the interfaces, they are nevertheless in line with typical literature values. \hl{It must of course be noted that a single cantilever cannot form a complete characterisation of an interfacial fracture toughness. As it is typically the Al$_2$O$_3$-Cr$_7$C$_3$ interface that fails to give alumina spallation at very high} temperatures~\cite{lee2014}, \hl{a complete characterisation of the Cr$_7$C$_3$-Cr$_2$AlC interface was not carried out.}

Spallation during thermal cycling comes from a build-up of thermal stresses during heating and cooling, as a result of differences in thermal expansion coefficients. From literature data, the coefficients of Al$_2$O$_3$, Cr$_3$C$_2$ (in lieu of Cr$_7$C$_3$) and Cr$_2$AlC are 9.3$\times 10^{-6}~K^{-1}$~\cite{Cao20041}, 10.4$\times 10^{-6}~K^{-1}$ ~\cite{Hussainova2006} and $\sim$12.2$\times 10^{-6}~K^{-1}$~\cite{Wang20141202}. For a 1000$^{\circ}$C increase in temperature, this amounts to approximately 1\% strain for Al$_2$O$_3$ and Cr$_3$C$_2$ upon heating to 1000$^{\circ}$C, rather below the 5\% failure strain seen in Figure~\ref{Beam1}. It is furthermore possible, given the fundamental coupling of elastic modulus and thermal expansion coefficient, that the similar moduli of the carbide and oxide layers - inclusive of the exact chemistry and defects - compared to the bulk MAX phase (Fig.~\ref{RoomTNano}) means they exhibit more similar coefficients of expansivity compared with the literature. Indeed, this same explanation has been proposed to explain the stability of thermal barrier coatings on Ti$_2$AlC~\cite{Smialek201677}. Additionally, the required crack path for the loss of the oxide and carbide layers would be extremely convoluted, due to the complex interface layout (Fig.~\ref{Microstructure}), thereby `artificially' increasing the fracture toughness of the system above the 1.62~MPa$\sqrt{m}$ of the weakest interface. As a result, the weakest point of Cr$_2$AlC materials at high temperature is the Al$_2$O$_3$/Cr$_7$C$_3$ interface, so a potential failure of a component will occur there. Indeed, the catastrophic failure at this interface has been already observed in Cr$_2$AlC/YSZ system under thermal shock conditions at 1300$^{\circ}$C after 268 hours~\cite{Gonzalez201817}, confirming the experimental values measured in this paper.

\section{Conclusions}
The MAX phase Cr$_2$AlC was observed to have excellent thermal stability, particularly in regards to burner rig conditions in which no spallation of oxide layers were seen even after 500 heating-cooling cycles due to the formation of a layer of Cr$_7$C$_3$ and Al$_2$O$_3$ on the surface.

Micro-cantilever fracture tests at these interfaces revealed typical fracture toughness values associated with ceramic materials, below 4~MPa$\sqrt{m}$, demonstrating that even after extreme environmental conditions, these interfaces exhibit typical ceramic properties, therefore implying no change in their mechanical properties. It is therefore hypothesised that the excellent macroscopic behaviour is due to a combination of low internal strains due to the similar coefficients of thermal expansion, and the convoluted interface structure preventing straight crack propagation.

The high-temperature mechanical properties of Cr$_2$AlC have also been studied by nanoindentation. These data show modest reductions in both hardness and modulus at temperatures up to 980$^{\circ}$C, implying no change in deformation mechanism up to these temperatures. The combination of oxidative testing and high-temperature nanoindentation therefore allows future MAX-phases or MAX-phase composite materials to be selected and rapidly evaluated based on a combination of their chemical and mechanical performance.

\section*{Funding}
Financial support from the JARA seed fund project `JARA-ENERGY002: Novel ceramic foams for heat exchangers at high temperature' is gratefully acknowledged.

\bibliographystyle{tfnlm}
\bibliography{JARAMAX}

\begin{thebibliography}{10}
\providecommand{\url}[1]{\normalfont{#1}}
\providecommand{\urlprefix}{Available from: }

\bibitem{Barsoum1999363}
Barsoum~MW, El-Raghy~T. Room-temperature ductile carbides. Metallurgical and
  Materials Transactions a-Physical Metallurgy and Materials Science.
  1999;\hspace{0pt}30(2):363--369.

\bibitem{Eklund20101851}
Eklund~P, Beckers~M, Jansson~U, et~al. The {M(n+1)AX(n)} phases: Materials
  science and thin-film processing. Thin Solid Films.
  2010;\hspace{0pt}518(8):1851--1878.

\bibitem{Barsoum2011195}
Barsoum~M, Radovic~M. Mechanical properties of the {MAX} phases. Vol.~41. ;
  2011.

\bibitem{Tian20071663}
Tian~W, Wang~P, Zhang~G, et~al. Mechanical properties of {Cr$_2$AlC} ceramics.
  Journal of the American Ceramic Society. 2007;\hspace{0pt}90(5):1663--1666.

\bibitem{Sun2011143}
Sun~ZM. Progress in research and development on {MAX} phases: a family of
  layered ternary compounds. International Materials Reviews.
  2011;\hspace{0pt}56(3):143--166.

\bibitem{Farle201537}
Farle~AS, Kwakernaak~C, van~der Zwaag~S, et~al. A conceptual study into the
  potential of {M(n+1)AX(n)}-phase ceramics for self-healing of crack damage.
  Journal of the European Ceramic Society. 2015;\hspace{0pt}35(1):37--45.

\bibitem{Li2013892}
Li~SB, Xiao~LO, Song~GM, et~al. Oxidation and crack healing behavior of a
  fine-grained {Cr$_2$AlC} ceramic. Journal of the American Ceramic Society.
  2013;\hspace{0pt}96(3):892--899.

\bibitem{Wang2010385}
Wang~XH, Zhou~YC. {Layered Machinable and Electrically Conductive Ti2AlC and
  Ti3AlC2 Ceramics: a Review}. Journal of Materials Science and Technology.
  {2010} {MAY};\hspace{0pt}{26}({5}):{385--416}.

\bibitem{Walter2006389}
Walter~C, Sigumonrong~DP, El-Raghy~T, et~al. {Towards large area deposition of
  Cr2AlC on steel}. Thin Solid Films. {2006} {OCT
  25};\hspace{0pt}{515}({2}):{389--393}. {12th International Conference on Thin
  Films, Bratislava, Slovakia, 2002}.

\bibitem{Shtansky20093595}
Shtansky~DV, Kiryukhantsev-Korneev~PV, Sheveyko~AN, et~al. {Comparative
  investigation of TiAlC(N), TiCrAlC(N), and CrAlC(N) coatings deposited by
  sputtering of MAX-phase Ti$_{2-x}$Cr$_x$AlC targets}. Surface and Coatings
  Technology. {2009} {AUG 25};\hspace{0pt}{203}({23}):{3595--3609}.

\bibitem{Horlait2016682}
Horlait~D, Grasso~S, Al~Nasiri~N, et~al. {Synthesis and Oxidation Testing of
  MAX Phase Composites in the Cr-Ti-Al-C Quaternary System}. Journal of the
  American Ceramic Society. {2016} {FEB};\hspace{0pt}{99}({2}):{682--690}.

\bibitem{Yu20105997}
Yu~WB, Li~SB, Sloof~WG. Microstructure and mechanical properties of a
  {Cr$_2$Al(Si)C} solid solution. Materials Science and Engineering
  a-Structural Materials Properties Microstructure and Processing.
  2010;\hspace{0pt}527(21-22):5997--6001.

\bibitem{Berger2017192}
Berger~O, Boucher~R. Crack healing in {Y}-doped {Cr$_2$AlC-MAX} phase coatings.
  Surface Engineering. 2017;\hspace{0pt}33(3):192--203.

\bibitem{Pei2017441}
Pei~R, McDonald~SA, Shen~L, et~al. Crack healing behaviour of {Cr$_2$AlC MAX}
  phase studied by x-ray tomography. Journal of the European Ceramic Society.
  2017;\hspace{0pt}37(2):441--450.

\bibitem{Shen20161}
Shen~L, Eichner~D, van~der Zwaag~S, et~al. Reducing the erosive wear rate of
  {Cr$_2$AlC MAX} phase ceramic by oxidative healing of local impact damage.
  Wear. 2016;\hspace{0pt}358-359:1--6.

\bibitem{Gupta2008560}
Gupta~S, Filimonov~D, Palanisamy~T, et~al. Tribological behavior of select
  {MAX} phases against {Al$_2$O$_3$} at elevated temperatures. Wear.
  2008;\hspace{0pt}265(3-4):560--565.

\bibitem{Gupta2008270}
Gupta~S, Filimonov~D, Zaitsev~V, et~al. Ambient and 550$^{\circ}${C}
  tribological behavior of select {MAX} phases against {Ni}-based superalloys.
  Wear. 2008;\hspace{0pt}264(3-4):270--278.

\bibitem{Li20132715}
Li~SB, Chen~XD, Zhou~Y, et~al. Influence of grain size on high temperature
  oxidation behavior of {Cr$_2$AlC} ceramics. Ceramics International.
  2013;\hspace{0pt}39(3):2715--2721.

\bibitem{Shang20171339}
Shang~L, Konda~Gokuldoss~P, Sandl\"{o}bes~S, et~al. Effect of {Si} additions on
  the {Al$_2$O$_3$} grain refinement upon oxidation of {Cr$_2$AlC MAX} phase.
  Journal of the European Ceramic Society. 2017;\hspace{0pt}37(4):1339--1347.

\bibitem{Berger2015373}
Berger~O, Boucher~R, Ruhnow~M. {Part I}. mechanism of oxidation of {Cr$_2$AlC}
  films in temperature range 700-1200 degrees {C}. Surface Engineering.
  2015;\hspace{0pt}31(5):373--385.

\bibitem{Yang2013203}
Yang~HJ, Pei~YT, De~Hosson~JTM. Oxide-scale growth on {Cr$_2$AlC} ceramic and
  its consequence for self-healing. Scripta Materialia.
  2013;\hspace{0pt}69(2):203--206.

\bibitem{Hajas2011591}
Hajas~DE, to~Baben~M, Hallstedt~B, et~al. {Oxidation of Cr$_2$AlC coatings in
  the temperature range of 1230 to 1410$^{\circ}$C}. Surface and Coatings
  Technology. 2011;\hspace{0pt}206(4):591--598.

\bibitem{Tallman2013115}
Tallman~DJ, Anasori~B, Barsoum~MW. A critical review of the oxidation of
  {Ti$_2$AlC, Ti$_3$AlC$_2$ and Cr$_2$AlC} in air. Materials Research Letters.
  2013;\hspace{0pt}1(3):115--125.

\bibitem{Smialek201723}
Smialek~JL. Environmental resistance of a {Ti$_2$AlC}-type {MAX} phase in a
  high pressure burner rig. Journal of the European Ceramic Society.
  2017;\hspace{0pt}37(1):23--34.

\bibitem{Gonzalez20181841}
Gonzalez-Julian~J, Go~T, Mack~DE, et~al. Environmental resistance of {Cr$_2$AlC
  MAX} phase under thermal gradient loading using a burner rig. Journal of the
  American Ceramic Society. 2018;\hspace{0pt}101(5):1841--1846.

\bibitem{Gonzalez201817}
Gonzalez-Julian~J, Go~T, Mack~DE, et~al. Thermal cycling testing of {TBC}s on
  {Cr$_2$AlC MAX} phase substrates. Surface and Coatings Technology.
  2018;\hspace{0pt}340:17--24.

\bibitem{Smialek201677}
Smialek~JL, Harder~BJ, Garg~A. Oxidative durability of {TBCs} on {Ti$_2$AlC}
  {MAX} phase substrates. Surface and Coatings Technology.
  2016;\hspace{0pt}285:77--86.

\bibitem{Smialek2018782}
Smialek~JL. Oxidation of {Al$_2$O$_3$} scale-forming {MAX} phases in turbine
  environments. Metallurgical and Materials Transactions a-Physical Metallurgy
  and Materials Science. 2018;\hspace{0pt}49A(3):782--792.

\bibitem{Gibson2017007}
Gibson~JSKL, Schr\"{o}ders~S, Zehnder~C, et~al. On extracting mechanical
  properties from nanoindentation at temperatures up to 1000$^{\circ}${C}.
  Extreme Mechanics Letters. 2017;\hspace{0pt}.

\bibitem{Korte2012167}
Korte~S, Stearn~RJ, Wheeler~JM, et~al. High temperature microcompression and
  nanoindentation in vacuum. Journal of Materials Research.
  2012;\hspace{0pt}27(1):167--176.

\bibitem{Harris20161}
Harris~AJ, Beake~BD, Armstrong~DEJ, et~al. Development of high temperature
  nanoindentation methodology and its application in the nanoindentation of
  polycrystalline tungsten in vacuum to 950$^{\circ}${C}. Experimental
  Mechanics. 2016;\hspace{0pt}:1--12.

\bibitem{Gonzalez2016415}
Gonzalez-Julian~J, Onrubia~S, Bram~M, et~al. Effect of sintering method on the
  microstructure of pure {Cr$_2$AlC MAX} phase ceramics. Journal of the Ceramic
  Society of Japan. 2016;\hspace{0pt}124(4):415--420.

\bibitem{Oliver19921564}
Oliver~W, Pharr~G. Improved technique for determining hardness and elastic
  modulus using load and displacement sensing indentation experiments. Journal
  of Materials Research. 1992;\hspace{0pt}7(6):1564--1580.

\bibitem{Oliver20043}
Oliver~WC, Pharr~GM. Measurement of hardness and elastic modulus by
  instrumented indentation: Advances in understanding and refinements to
  methodology. Journal of Materials Research. 2004;\hspace{0pt}19(1):3--20.

\bibitem{Maio2005299}
Di~Maio~D, Roberts~SG. Measuring fracture toughness of coatings using
  focused-ion-beam-machined microbeams. Journal of Materials Research.
  2005;\hspace{0pt}20(2):299--302.

\bibitem{Naveed201611}
Naveed~M, Obrosov~A, Zak~A, et~al. Sputtering power effects on growth and
  mechanical properties of {Cr$_2$AlC} {MAX} phase coatings. Metals.
  2016;\hspace{0pt}6(11):11.

\bibitem{Sun2004589}
Sun~Z, Li~S, Ahuja~R, et~al. Calculated elastic properties of {M2AlC (M=Ti, V,
  Cr, Nb and Ta)}. Solid State Communications.
  2004;\hspace{0pt}129(9):589--592.

\bibitem{Hou20051039}
Hou~QR, Zhang~HY, Chen~YB. Deposition of chromium-carbon films by magnetron
  sputtering of chromium and carbon targets. Modern Physics Letters B.
  2005;\hspace{0pt}19(21):1039--1050.

\bibitem{Paul2002123}
Paul~A, Lim~J, Choi~K, et~al. Effects of deposition parameters on the
  properties of chromium carbide coatings deposited onto steel by sputtering.
  Materials Science and Engineering: {A}. 2002;\hspace{0pt}332(1):123--128.

\bibitem{Xiao2009115415}
Xiao~B, Xing~JD, Feng~J, et~al. A comparative study of {Cr$_7$C$_3$ , Fe$_3$C
  and Fe$_2$B} in cast iron both from ab initio calculations and experiments.
  Journal of Physics D: Applied Physics. 2009;\hspace{0pt}42(11):115415.

\bibitem{Hussainova2006}
Hussainova~I, Antonov~M. Thermophysical properties and thermal shock resistance
  of chromium carbide based cermets. Vol.~12. ; 2006.

\bibitem{Auerkari1996}
Auerkari~P. Mechanical and physical properties of engineering alumina ceramics.
  Vol.~23. ; 1996.

\bibitem{Lapauw20165445}
Lapauw~T, Tytko~D, Vanmeensel~K, et~al. {(Nb$_x$, Zr$_{1-x}$)$_4$AlC$_3$ MAX}
  phase solid solutions: Processing, mechanical properties, and density
  functional theory calculations. Inorganic Chemistry.
  2016;\hspace{0pt}55(11):5445--5452.

\bibitem{Khatun20173634}
Khatun~MR, Ali~MA, Parvin~F, et~al. Elastic, thermodynamic and optical behavior
  of {V$_2$AC (A=Al, Ga) MAX} phases. Results in Physics.
  2017;\hspace{0pt}7:3634--3639.

\bibitem{Ali20124221}
Ali~MS, Islam~A, Hossain~MM, et~al. Phase stability, elastic, electronic,
  thermal and optical properties of {Ti$_3$Al$_{1-x}$Si$_x$C$_2$ (0 $<=$ x $<=$
  1)}: First principle study. Physica B-Condensed Matter.
  2012;\hspace{0pt}407(21):4221--4228.

\bibitem{Bhattacharya2014319}
Bhattacharya~R, Benitez~R, Radovic~M, et~al. High strain-rate response and
  deformation mechanisms in polycrystalline {Ti$_2$AlC}. Materials Science and
  Engineering a-Structural Materials Properties Microstructure and Processing.
  2014;\hspace{0pt}598:319--326.

\bibitem{Guitton20146358}
Guitton~A, Joulain~A, Thilly~L, et~al. Evidence of dislocation cross-slip in
  {MAX} phase deformed at high temperature. Scientific Reports.
  2014;\hspace{0pt}4:6358.

\bibitem{ROA2018211}
Roa~J, Phani~PS, Oliver~W, et~al. Mapping of mechanical properties at
  microstructural length scale in wc-co cemented carbides: Assessment of
  hardness and elastic modulus by means of high speed massive nanoindentation
  and statistical analysis. International Journal of Refractory Metals and Hard
  Materials. 2018;\hspace{0pt}75:211 -- 217.
  \urlprefix\url{http://www.sciencedirect.com/science/article/pii/S0263436817308995}.

\bibitem{Wang20182073}
Wang~NC, Jiang~F, Xu~XP, et~al. Effects of crystal orientation on the crack
  propagation of sapphire by sequential indentation testing. Crystals.
  2018;\hspace{0pt}8(1).

\bibitem{Huang201107}
Huang~JL, Nayak~P. Processing and characterization of alumina/ chromium carbide
  ceramic nanocomposite. Rijeka: InTech; 2011. p. Ch. 07.

\bibitem{lee2014}
Lee~DB. Critical review of the oxidation of cr2alc. In: 13th International
  Ceramics Congress - Part C; (Advances in Science and Technology; Vol.~89);
  12. Trans Tech Publications Ltd; 2014. p. 115--122.

\bibitem{Cao20041}
Cao~XQ, Vassen~R, Stoever~D. Ceramic materials for thermal barrier coatings.
  Journal of the European Ceramic Society. 2004;\hspace{0pt}24(1):1--10.

\bibitem{Wang20141202}
Wang~J, Wang~J, Li~A, et~al. Theoretical study on the mechanism of anisotropic
  thermal properties of {Ti$_2$AlC} and {Cr$_2$AlC}. Journal of the American
  Ceramic Society. 2014;\hspace{0pt}97(4):1202--1208.

\end{thebibliography}

\end{document}